\documentclass[aps,prl,reprint]{revtex4-1}

\usepackage{graphicx,graphics,amsfonts,amsmath,amssymb}

\usepackage{color}
\usepackage{verbatim}
\usepackage{tikz}

\begin{document}



\title{Ultracold Dipolar Gas of Fermionic $^{23}$Na$^{40}$K Molecules in their Absolute Ground State}

\author{Jee Woo Park, Sebastian A. Will, and Martin W. Zwierlein}
    \affiliation{MIT-Harvard Center for Ultracold Atoms, Research Laboratory of Electronics, and Department of Physics, Massachusetts Institute of Technology,
    Cambridge, Massachusetts 02139, USA }

\date{\today}

\begin{abstract}
We report on the creation of an ultracold dipolar gas of fermionic $^{23}$Na$^{40}$K molecules in their absolute rovibrational and hyperfine ground state. Starting from weakly bound Feshbach molecules, we demonstrate hyperfine resolved two-photon transfer into the singlet ${\rm X}^1\Sigma^+ |v{=}0,J{=}0\rangle$ ground state, coherently bridging a binding energy difference of 0.65 eV via stimulated rapid adiabatic passage. The spin-polarized, nearly quantum degenerate molecular gas displays a lifetime longer than 2.5 s, highlighting NaK's stability against two-body chemical reactions. A homogeneous electric field is applied to induce a  dipole moment of up to 0.8 Debye. With these advances, the exploration of many-body physics with strongly dipolar Fermi gases of $^{23}$Na$^{40}$K molecules is in experimental reach.
\end{abstract}

\maketitle


Quantum gases of dipolar molecules promise to become a platform for precision measurements, quantum information processing, high-speed quantum simulation and the creation of novel many-body systems~\cite{krem09coldmolecules,Carr09mol, Baranov2012}. The long-range, anisotropic nature of electric dipolar interactions between molecules is expected to yield novel types of order, such as topological superfluidity with fermionic molecules~\cite{Baranov2002,Cooper:2009, Levinsen:2011}, interlayer pairing between two-dimensional systems~\cite{pikovski2010bilayer,Baranov:2011}, and the formation of dipolar quantum crystals~\cite{Sieberer:2011}.
The necessary prerequisite is the full control over the molecules' translational, electronic, vibrational, rotational and nuclear spin degrees of freedom~\cite{Quemener:2012}. In pioneering work on $^{40}$K$^{87}$Rb and non-dipolar $^{133}$Cs$_2$, weakly bound molecules were associated from ultracold atoms via Feshbach resonances, and subsequently coherently transferred into the absolute rovibrational ground state via a two-photon stimulated rapid adiabatic passage (STIRAP)~\cite{Ospelkaus:2008,ni08polar,Danzl2010groundstate,danzl08Csmol,Aikawa:2010}. Hyperfine control of the KRb molecules was demonstrated using microwave radiation~\cite{OspelkausS:2010}. Ground state KRb molecules are chemically unstable against two-body collisions, as the reaction $\rm{KRb}+\rm{KRb}\rightarrow\rm{K}_{2}+\rm{Rb}_{2}$ is energetically allowed. This enabled studies of quantum-state controlled chemical reactions~\cite{ospe10chemical,Ni2010dipolar}, but also led to loss and heating of the trapped molecules. Loading of KRb molecules into optical lattice potentials efficiently suppressed the reactions and enhanced the lifetime of molecular samples~\cite{deMiranda:2011,Chotia:2012,Yan:2013}.

\begin{figure}[h!]
  \begin{center}
  \includegraphics[width=.96\columnwidth]{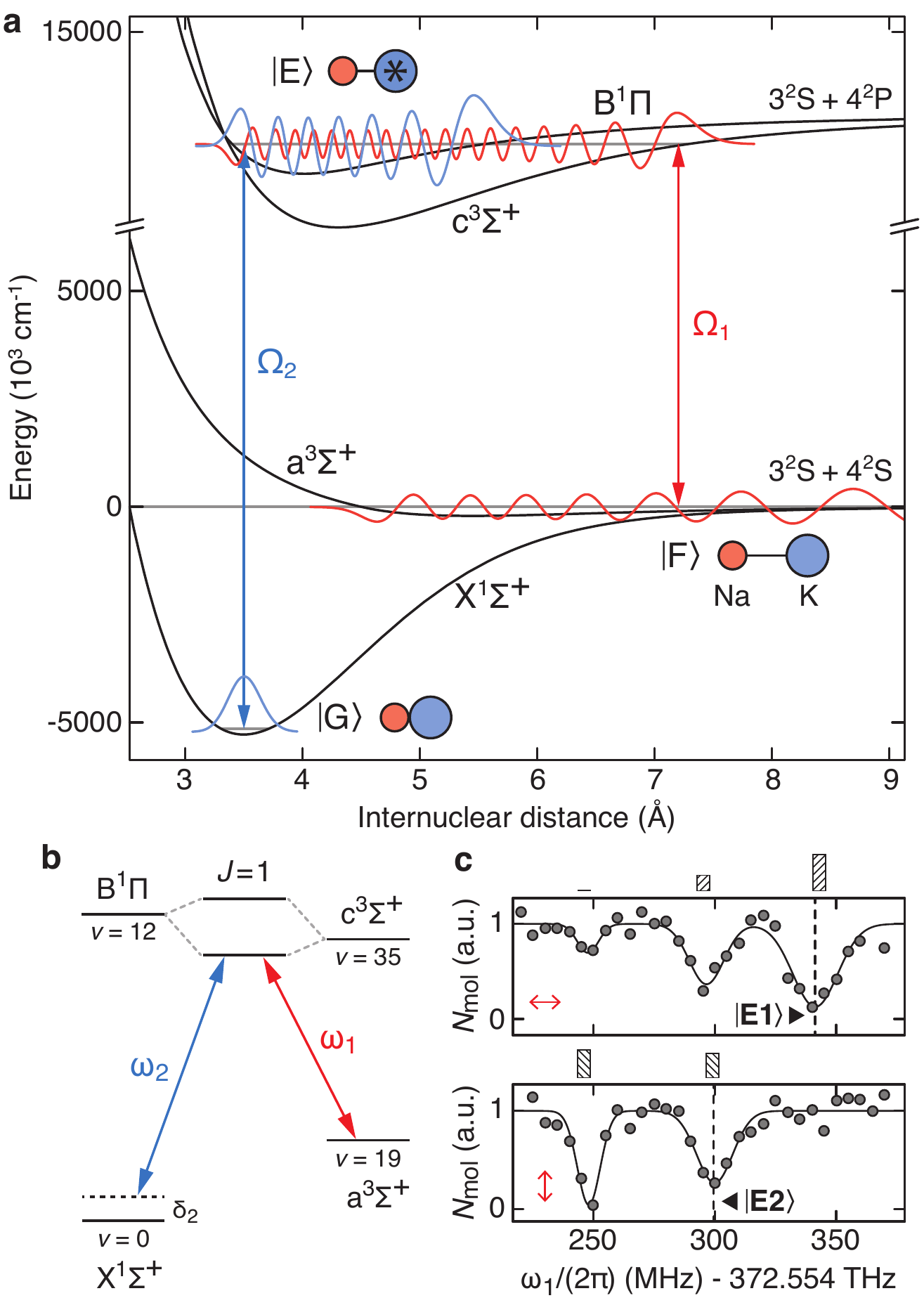}
  \end{center}
  \caption{\label{fig:cartoon} Two-photon pathway to the absolute ground state of $^{23}$Na$^{40}$K. (a) Relevant molecular potentials for the STIRAP transfer to the rovibrational ground state $|v{=}0, J{=}0\rangle$ in the ${\rm X}^{1}\Sigma^+$ potential. Rabi frequencies of the Raman lasers are denoted by $\Omega_1$ and $\Omega_2$. (b) The initial Fesh\-bach molecular state is associated with the highest vibrational state $|v{=}19\rangle$ of the ${\rm a}^{3}\Sigma^+$ potential. In the excited state, spin-orbit coupling resonantly mixes the vibrational states ${\rm c}^{3}\Sigma^+|v{=}35, J{=}1\rangle$ and ${\rm B}^{1}\Pi|v{=}12, J{=}1\rangle$ with 64\% triplet and 36\% singlet character~\cite{Park2015:1}. $\delta_2$ denotes the two-photon detuning of the Raman lasers. (c) Single-photon spectra via laser 1 resolving hyperfine states $|{\rm E}1\rangle$ and $|{\rm E2}\rangle$ of the electronically excited state. In the upper (lower) panel, the linear polarization of laser 1 is chosen to be horizontal (vertical), which is perpendicular (parallel) to the quantization axis set by a vertical magnetic field. The bars above the upper (lower) panel indicate the expected relative spectral weight of the contributing hyperfine states with $m_F {=-}5/2$ ($m_F {=-} 7/2$)~\cite{Park2015:1}.
  }
\end{figure}

For the study of collisionally dense molecular gases in the quantum regime, molecules that are stable against two-body collisions are of great interest. Possible choices among the bialkalis were summarized in~\cite{zuch10mol}, for example RbCs and NaK. Recently, ultracold bosonic RbCs molecules were successfully coherently transferred into the rovibrational ground state~\cite{Takekoshi2014RbCs,Molony2014}.
For fermionic molecules, the only chemically stable bialkalis are $^{23}$Na$^{40}$K and $^{40}$K$^{133}$Cs, with NaK possessing the larger electric dipole moment of $d=2.72$ Debye~\cite{worm81nak, gerd11nak}.
Feshbach molecules of $^{23}$Na$^{40}$K have been produced in earlier work by our group~\cite{Park:2012,wu2012NaK}.
Here, we report on the creation of an ultracold dipolar gas of fermionic $^{23}$Na$^{40}$K ground state molecules. Using STIRAP we demonstrate efficient transfer between the Feshbach molecular state and the rovibrational ground state, coherently bridging the molecular binding energy difference of 0.65 eV. Applying a static electric field, Stark shifts of the molecular ground state of up to 400 MHz are observed, corresponding to an induced dipole moment of $d=0.8$ Debye, significantly larger than the maximal dipole moment achieved in previous studies~\cite{Molony2014}. At $d=0.8$~Debye, the characteristic range of dipolar interactions $a_{\rm d} = m_{\rm NaK} d^2/(4 \pi \epsilon_0 \hbar^2) $ ($m_{\rm NaK}$ denotes the molecular mass) reaches $0.6$~$\mu$m, comparable to the interparticle spacing of $1.6$~$\mu$m for the realized peak densities of $n_0 = 2.5 \times 10^{11}$~cm$^{-3}$. The corresponding dipolar interaction energy $E_{\rm d} = d^2 n_0 /(4\pi \epsilon_0)$ approaches 5\% of the local Fermi energy, and should therefore dominate the many-body physics of the gas.


Our starting point is a gas of about $7\times 10^3$ Feshbach molecules of $^{23}$Na$^{40}$K, trapped in a crossed optical dipole trap at a temperature of $ T \approx 500\,\rm nK$, corresponding to $T/T_{\rm F,mol} \approx 2.0$, where $T_{\rm F, mol}$ is the Fermi temperature of trapped molecules~\cite{Park:2012,wu2012NaK}\footnote{The trap frequencies for $^{40}$K [$^{23}$Na] are $\nu_x = 190(15)$ Hz [$150(10)$ Hz], $\nu_y = 160(15)$ Hz [$120(10)$ Hz], and $\nu_z = 130(15)$ Hz [$110(10)$ Hz].}. The molecules are created via radiofrequency association at a field of 85.7~G, near an $s$-wave Fesh\-bach resonance that couples pairs of Na and K atoms in their lowest hyperfine states $|f_{\rm Na}, m_{f_{\rm Na}}\rangle= |1, 1\rangle$ and $|f_{\rm K}, m_{f_{\rm K}}\rangle= |9/2, -9/2\rangle$ to the triplet molecular state ${\rm a}^3\Sigma^+ |v{=}19,J{=}1,F{=}9/2,m_F{=-}7/2\rangle$~\cite{Park:2012}. Here, the quantum numbers $J$ and $F$ refer to the molecule's total angular momentum $\vec{J}$ and $\vec{F}$ excluding and including nuclear spins, respectively, $m_F$ describes the projection of $\vec{F}$ along the quantization axis, and $v$ denotes the vibrational state.
The molecular binding energy is $h \times 80(5)\,\rm kHz$, small enough to allow for direct absorption imaging of Feshbach molecules with light resonant with the atomic cycling transition of $^{40}$K~\cite{wu2012NaK}.


\begin{figure}
  \begin{center}
  \includegraphics[width=.96\columnwidth]{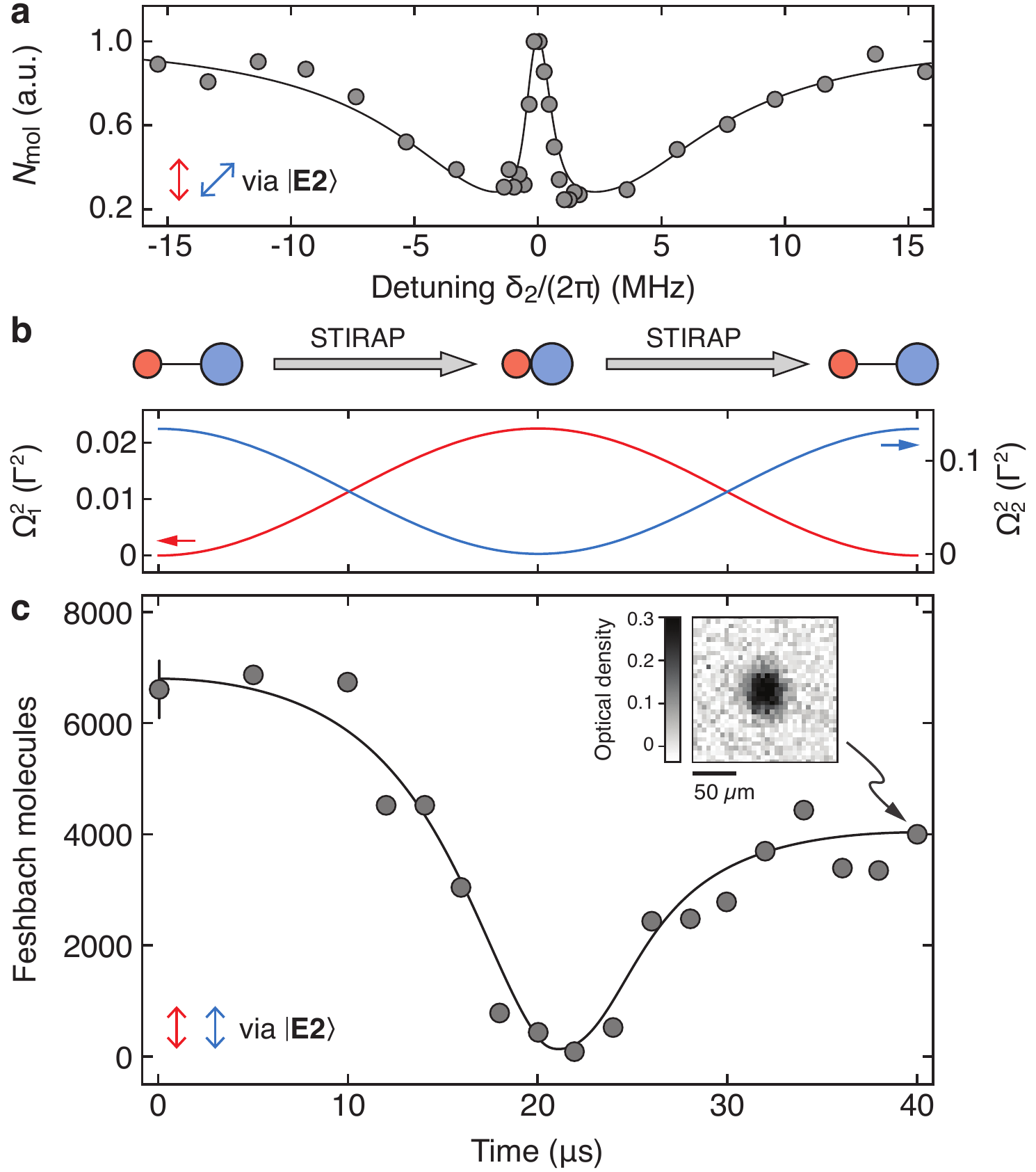}
  \end{center}
  \caption{\label{fig:STIRAP} Creation of $^{23}$Na$^{40}$K ground state molecules. (a) Observation of the rovibrational ground state using two-photon dark resonance spectroscopy. The solid line is obtained by fitting a three-level master equation model using an excited state linewidth of $\Gamma = 2 \pi \times 9$ MHz. (b) The STIRAP sequence realizes coherent transfer from the Feshbach state to the rovibrational ground state and back via sinusoidal ramping of the laser intensities. (c) Evolution of the number of Feshbach molecules during the STIRAP sequence. The solid line is obtained from the master equation model using the Rabi couplings $\Omega_{1}/\Gamma=0.15$ and $\Omega_{2}/\Gamma=  0.37$.  The inset shows a background free absorption image of about $4 \times 10^3$ Feshbach molecules after a STIRAP round trip.}
\end{figure}

Raman lasers connect the initial Feshbach molecular state with dominant spin triplet character to the rovibrational singlet ground state via an excited state with strongly mixed singlet and triplet spin character (see Fig.~\ref{fig:cartoon})~\cite{Stwalley:2004,Schulze:2013,Park2015:1}. In contrast to KRb~\cite{ni08polar} and RbCs~\cite{Sage:2005} where a range of electronically excited vibrational states have a mixed spin character, NaK offers only few such states due to comparatively weak spin-orbit coupling~\cite{Ferber2000NaK, Schulze:2013, Park2015:1}. In this work we use two specific hyperfine states of the resonantly mixed complex $\mathrm{c}^{3}\Sigma^+|v{=}35, J{=}1\rangle{\sim} \mathrm{B}^{1}\Pi|v{=}12, J{=}1\rangle$, labelled $|\rm{E}1\rangle$ and $|\rm{E2}\rangle$ in Fig.~\ref{fig:cartoon}(c), which give access to different hyperfine states in the rovibrational ground state \footnote{The state {$|\rm{E1}\rangle$} is described by quantum numbers $m_F{=-}5/2$, $F_1 {\approx}\, 5/2$, $m_{F_1}{\approx}\, 3/2$, while {$|\rm{E2}\rangle$} has $m_F{=-}7/2$, $F_1 {\approx}\, 5/2$, $m_{F_1}{\approx}\, 1/2$. Here, $F_1$ and $m_{F_1}$ refer to the total angular momentum excluding the $^{40}$K nuclear spin and its $z$-projection~\cite{Ishikawa1992c3shyperfine,Park2015:1}.}.
The detuning between the two Raman lasers is determined by the binding energy of the absolute ground state ${\rm X}^1\Sigma^+ |v{=}0,J{=}0\rangle$. We measure the binding energy by performing dark resonance spectroscopy [as in Fig.~\ref{fig:STIRAP}(a)], precisely calibrating the laser frequencies $\omega_1$ and  $\omega_2$ with an optical frequency comb and iodine spectroscopy, respectively. We obtain a value of  $D_0^{\rm(X)} = h \times 156.253169(3)$~THz~\cite{Park2015:1}, relative to the hyperfine center of mass of the constituent atoms. This represents a thousand-fold improvement in accuracy compared to previous determinations~\cite{gerd08nak}.

\begin{figure}
  \begin{center}
  \includegraphics[width=.96\columnwidth]{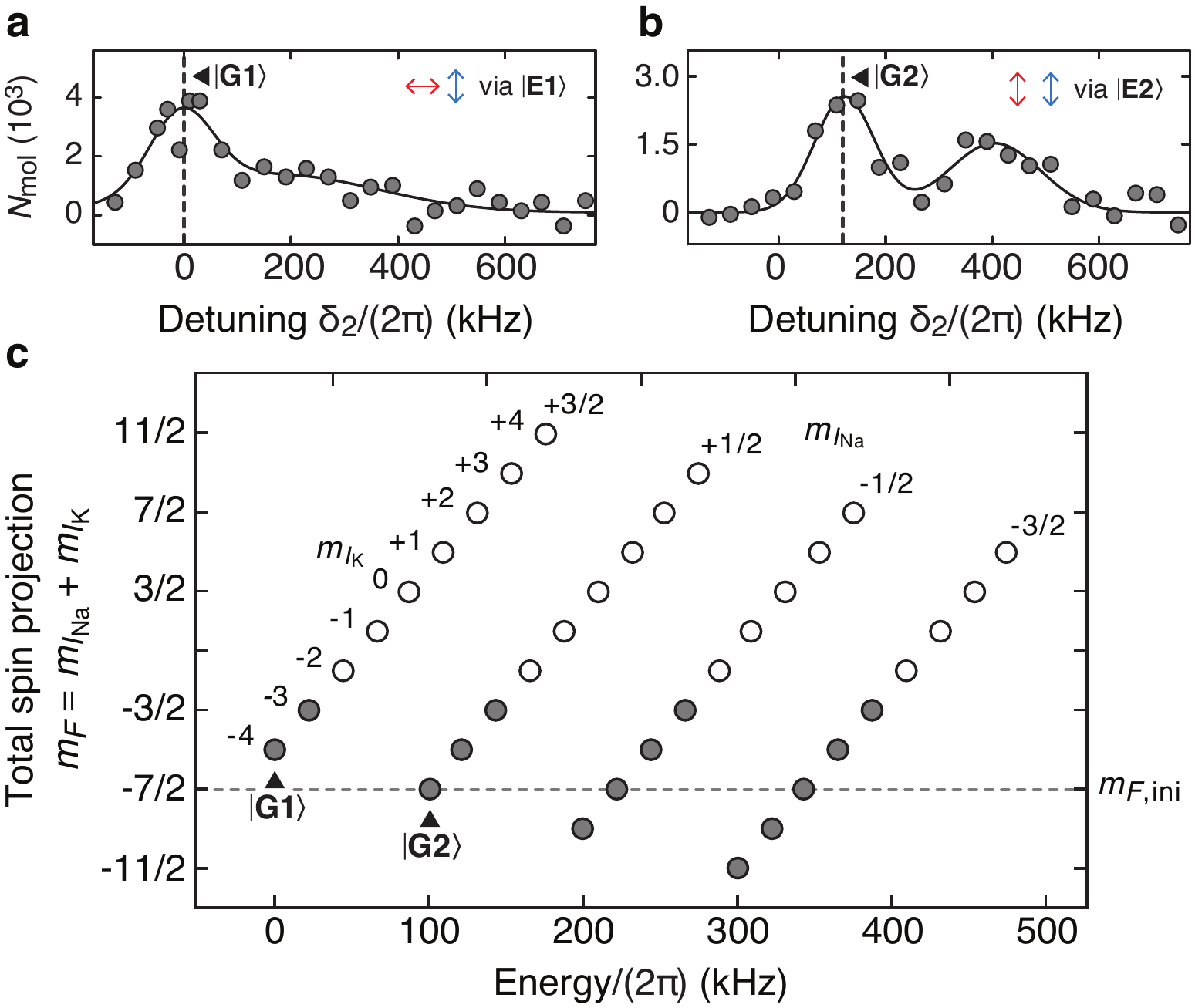}
  \end{center}
  \caption{\label{fig:hyperfine}
  Hyperfine structure of the rovibrational ground state of $^{23}$Na$^{40}$K. Hyperfine states $|{\rm G1}\rangle$ and $|{\rm G2}\rangle$ are detected using STIRAP resonances via intermediate states $|{\rm E}1\rangle$ (a) and $|{\rm E}2\rangle$ (b). The ground hyperfine state $|{\rm G}1\rangle$ is observed via $|{\rm E}1\rangle$ using horizontal and vertical polarization for laser 1 and 2, respectively, while $|{\rm G}2\rangle$ is accessed with both laser 1 and 2 vertically polarized. Solid lines are a guide to the eye. (c) Energies of all $(2 I_{\rm Na} +1)( 2 I_{\rm K}+1){=}36$ ground state hyperfine states at a magnetic field of 85.7 G. The nuclear Zeeman effect dominates, with weak influence of the scalar spin-spin interaction $c_4 \vec{I}_{\rm Na} \cdot \vec{I}_{\rm K}$ with a predicted $c_4{=-}466.2$ Hz~\cite{Aldegunde:2008, Hutson:2015}. The initial Feshbach state has $m_{F,\rm ini}{=-}7/2$. Hyperfine states that can be addressed in principle from this state using a two-photon process are shown as filled dots.}
\end{figure}

A crucial requirement for efficient STIRAP is a high degree of phase coherence between the Raman lasers. In our setup, a titanium-sapphire laser (laser 1 in Fig.~\ref{fig:cartoon}) couples the Feshbach molecular state to the excited state at a wavelength of 804.7~nm; that state in turn is coupled to the absolute singlet ground state with a dye laser (laser 2) using Rhodamine 6G dye at 566.9~nm. In order to maintain phase coherence, the Raman lasers are locked to a passively stable optical cavity made from ultralow expansion glass. At both wavelengths, the finesse of the cavity is about 15,000 and laser line widths on the kilohertz level are achieved via the Pound-Drever-Hall technique. For the dye laser, fast frequency correction is achieved via an intracavity electro-optic modulator~\cite{Kirchmair:2006}.

Figure~\ref{fig:STIRAP}(c) shows the STIRAP transfer of $^{23}$Na$^{40}$K Fesh\-bach molecules into the singlet rovibrational ground state, followed by the reverse transfer back into Feshbach molecules. The intensities of the lasers (proportional to $\Omega_{1,2}^2$) are ramped sinusoidally to adiabatically convert the dark state $\propto \Omega_2 | \rm{F} \rangle -\Omega_1 | \rm{G} \rangle$ of the coupled three-level system from the Feshbach state $| \rm{F} \rangle$ into the rovibrational ground state $| \rm{G} \rangle$ and back~\cite{Bergmann:1998}. We obtain a single-pass conversion efficiency of 75\%. About $5 \times 10^3$ ground state molecules are created; based on the in-situ size of the molecular cloud we estimate a peak density of $n_0 = 2.5 \times 10^{11}$ cm$^{-3}$. In order to record background free images of molecules, we perform STIRAP into the ground state and subsequently remove all remaining atoms and Feshbach molecules with resonant laser light. An exemplary image of a molecular cloud after a STIRAP round trip is shown in the inset of Fig.~\ref{fig:STIRAP}(c).


\begin{figure}
  \begin{center}
  \includegraphics[width=.96\columnwidth]{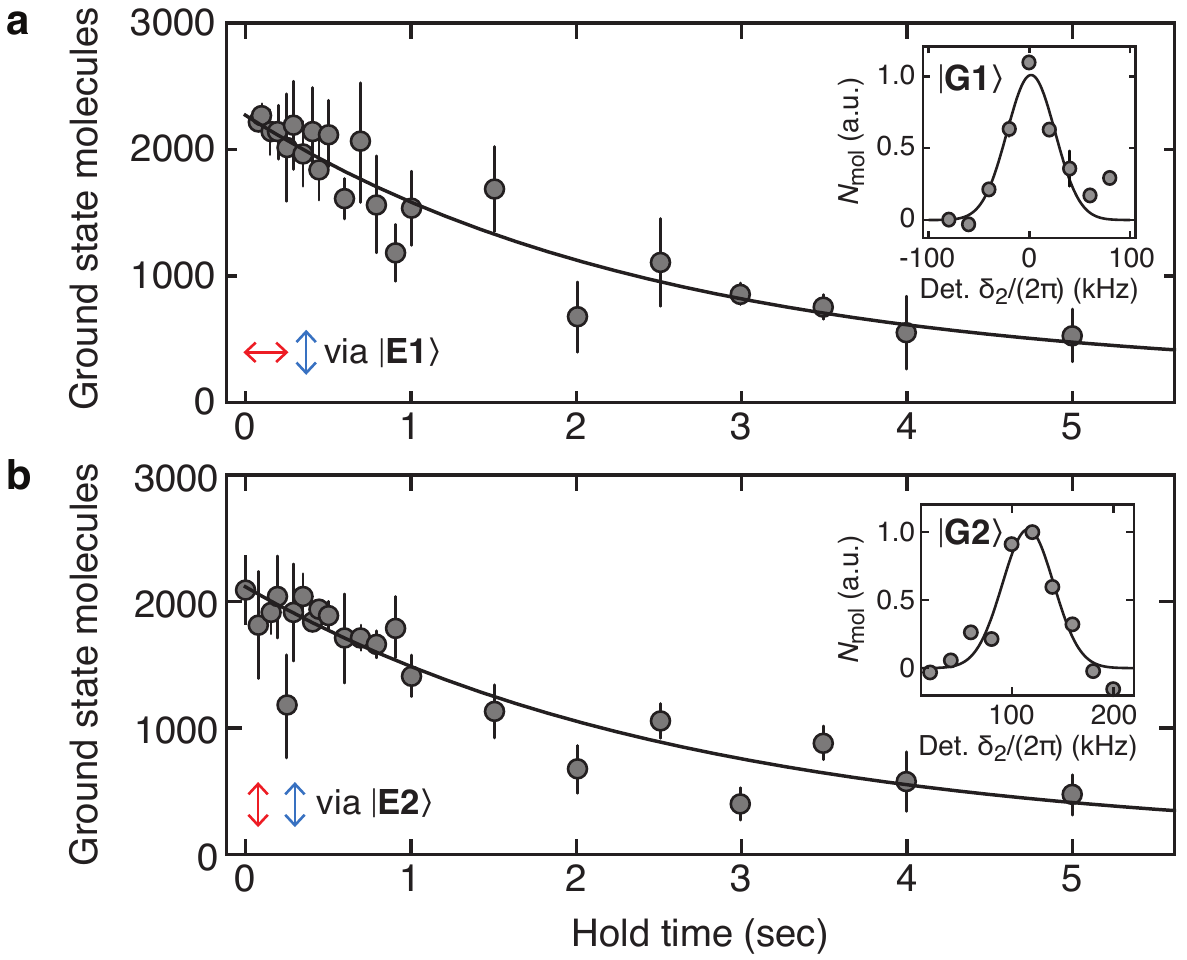}
  \end{center}
  \caption{\label{fig:lifetime} Lifetime of $^{23}$Na$^{40}$K in the rovibrational ground state. (a) Molecules in the absolute lowest hyperfine state $|{\rm G}1\rangle$ are created using STIRAP transfer via $|{\rm E}1\rangle$ and (b) in the absolute lowest hyperfine state $|{\rm G}2\rangle$ via $|{\rm E}2 \rangle$. Solid lines show exponential fits $A + B e^{-t/\tau}$ and yield lifetimes $\tau$ of 2.5(5)~s and 2.7(8)~s, respectively. Data points show the average of three experimental runs; the error bars denote the standard deviation of the mean. The insets show the respective STIRAP resonances, sufficiently narrow to address individual hyperfine states.}
\end{figure}

Controlled population of ground state hyperfine levels is achieved by choosing distinct intermediate excited states and laser polarizations (see Fig.~\ref{fig:hyperfine}). With the quantization axis defined by the vertical magnetic field, polarizations are chosen vertical or horizontal, corresponding to $\pi$ or $(\sigma^+ + \sigma^-)/\sqrt{2}$ light, respectively. The initial Feshbach state has total angular momentum projection $m_F {=-}7/2$. Accordingly, horizontal polarization of laser~1 and vertical polarization of laser~2 can access ground state hyperfine levels with $m_F {=-}9/2$ and $-5/2$. Using the intermediate state $|\rm{E1}\rangle$, which has $m_F {=-}5/2$, we observe dominant STIRAP transfer into the lowest hyperfine state $|\rm{G1}\rangle$ [see Fig.~\ref{fig:hyperfine}(a)]. 
Similarly, vertical polarization for both laser 1 and 2 leaves $m_F {=-}7/2$ unchanged and allows dominant coupling into the excited hyperfine state $|\rm{G2}\rangle$ via the electronically excited state $|\rm{E2}\rangle$ [see Fig.~\ref{fig:hyperfine}(b)]. 
From the experimental data we obtain a splitting between $|\rm{G1}\rangle$ and $|\rm{G2}\rangle$ of 120(30) kHz, in agreement with the prediction based on the dominant nuclear Zeeman effect and the weak scalar nuclear spin-spin interaction [see Fig.~\ref{fig:hyperfine}(c)].

\begin{figure}
  \begin{center}
  \includegraphics[width=.96\columnwidth]{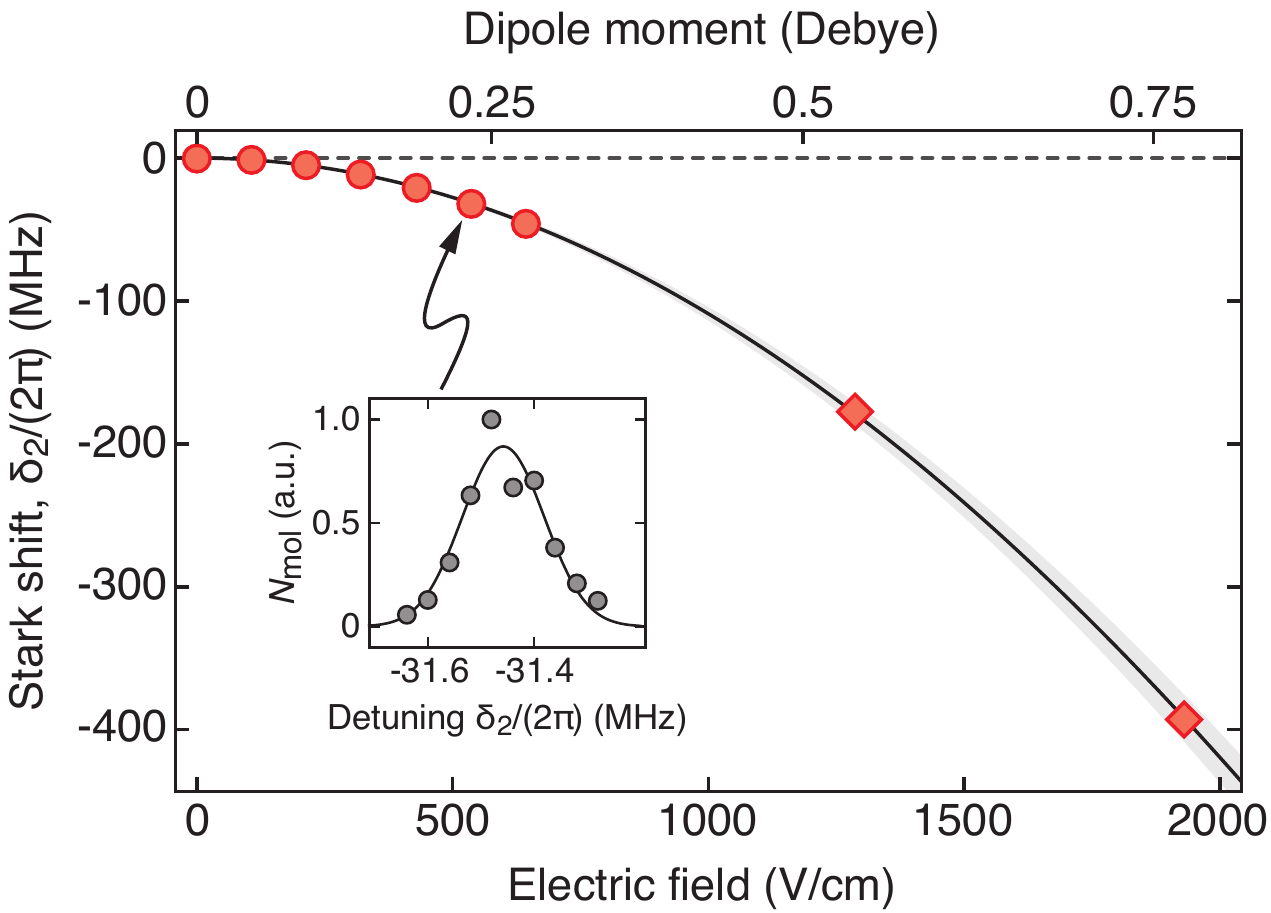}
  \end{center}
  \caption{\label{fig:Stark} Creation of a strongly dipolar gas of $^{23}$Na$^{40}$K molecules. The Stark shift in the presence of a homogeneous electric field is measured via STIRAP resonances (circles) and dark resonance spectroscopy (squares). Error bars are smaller than the point size. In a STIRAP transfer at finite electric field strongly dipolar ground state molecules are created (see inset). The solid line corresponds to the Stark shift expected for the known permanent dipole moment $d$ and the rotational constant $B$ of the ground state. The gray shading reflects the uncertainty. The Stark shift is used to calibrate the electric field strength.}
\end{figure}

$^{23}$Na$^{40}$K presents us with the opportunity to study the collisional properties of a gas of fermionic molecules that are chemically stable against two-body collisions~\cite{zuch10mol}. To this end, we measure the lifetime of $^{23}$Na$^{40}$K molecules in the absolute lowest hyperfine state $|\rm{G1}\rangle$ as well as in the hyperfine state $|\rm{G2}\rangle$ by varying the hold time between the STIRAP transfer into and out of these states (see Fig.~\ref{fig:lifetime}). To ensure a high degree of purity in the preparation of individual hyperfine states, the width of the STIRAP resonance is narrowed to less than 100 kHz (see insets of Fig.~\ref{fig:lifetime}). In each hyperfine state $|\rm{G1}\rangle$ and $|\rm{G2}\rangle$ $1/e$-lifetimes of about 2.5 s are observed. It is interesting to compare our data with a decay model that assumes unity loss for every short-range encounter of two molecules, as would be the case for chemically reactive molecules~\cite{ospe10chemical,Quemener:2010}. Spin-polarized fermionic molecules that dominantly undergo $p$-wave collisions need to tunnel through a $p$-wave barrier to meet at short-range. The temperature-dependent rate of close-range encounters is given by $\beta \propto T/E_p^{3/2}$~\cite{Quemener:2010}, where $E_p \propto (m_{\rm NaK}^3 C_6)^{-1/2}$ is the height of the barrier.
For NaK ground state molecules, the $C_6$ coefficient is dominated by virtual transitions to the nearby rotational states, $C_6 \approx d^4/(6 B) = 5.1(5) \times 10^5$ a.u.~\cite{Zuch2013,Lepers2013}, with the rotational constant $B=2.83\,\rm GHz$~\cite{Russier2000NaK}. For our temperature of $T=500 \, \rm nK$ and the barrier height of $E_p=12\,\rm \mu K$, we find $\beta \approx 6 \times 10^{-11}\,\rm cm^{-3}/s$ and, taking the average density of $\bar{n} =  0.4 \times 10^{11}\, \rm cm^{-3}$, a characteristic time scale for the decrease of the total molecule number of $1/(\beta \bar{n}) = 0.4$~s \footnote{Two-body decay of the molecular density follows the differential equation $\partial_{t} n {=-} \beta n^2 $. As the cloud has a Gaussian density distribution of fixed width, i.e. constant temperature, the total molecule number decays as $N(t) = N_0/(1+\beta \bar{n} t)$, where $N_0$ is the initial total number of molecules, and $\bar{n}=n_0/2^{3/2}$.}. By fitting a two-body decay curve to the data, one obtains a significantly slower characteristic time scale of 1.7~s, indicating an enhanced stability due to the absence of two-body chemical reactions.
At this point, technical reasons, such as imperfect polarization of the Raman lasers, off-resonant light scattering or collisions with a small number of remaining atoms, may still constitute a limit to the observed lifetime, in addition to potentially enhanced three-body loss from long-lived binary complexes~\cite{Mayle2013sticky}. The nature of the molecular loss process will be investigated in future work.
When ground state molecules are not prepared in a pure hyperfine state, but in a mixture of states due to Fourier or power broadening of the STIRAP resonance, lifetimes are significantly reduced by typically more than an order of magnitude. The much longer lifetimes in spin-polarized samples are likely due to the suppression of $s$-wave collisions as a consequence of the Pauli principle~\cite{ospe10chemical,Ni2010dipolar}.


$^{23}$Na$^{40}$K ground state molecules can interact via long-range interactions when their dipole moments are aligned in the laboratory frame. We demonstrate the alignment of the molecular dipoles by measuring the Stark shift of the ground state in the presence of a homogeneous electric field (see Fig.~\ref{fig:Stark}). Electric field strengths of up to 2~kV/cm are reached at the location of the molecules using indium tin oxide coated glass electrodes that are positioned outside of the ultra-high vacuum chamber. For field strengths up to 0.6~kV/cm, we use the Stark shift of the STIRAP resonance to characterize the influence of the electric field. The STIRAP transfer creates molecular samples with dipole moments of up to 0.3 Debye. For larger fields,  the Stark effect is observed in two-photon dark resonance spectra. At the maximal field, a Stark shift of about 400 MHz is observed, corresponding to a dipole moment of about $d= 0.8$ Debye.


In conclusion, we have created an ultracold gas of dipolar fermionic $^{23}$Na$^{40}$K molecules. Using STIRAP we have demonstrated comprehensive control over all molecular quantum states, preparing the molecules in their absolute electronic, vibrational, rotational and hyperfine ground state. In particular, following the preparation in a defined hyperfine state, lifetimes longer than 2.5 s are observed. By applying STIRAP in the presence of strong  electric fields we have created molecular samples with dipole moments of up to 0.3 Debye. Employing the tunability and strength of the electric dipole moment, ultracold samples of $^{23}$Na$^{40}$K should allow entering the strongly interacting regime of dipolar quantum matter, raising prospects for the creation of novel quantum phases.

We would like to thank Eberhard Tiemann, Thomas Bergeman, Amanda Ross, Robert Field, Rainer Blatt, Marco Prevedelli, Kang-Kuen Ni and Jun Ye for fruitful discussions and Jennifer Schloss, Emilio Pace, Huanqian Loh, and Zoe Yan for experimental assistance. This work was supported by the NSF, AFOSR PECASE, ARO, an ARO MURI on ``High-Resolution Quantum Control of Chemical Reactions'', an AFOSR MURI on ``Exotic Phases of Matter'', and the David and Lucile Packard Foundation.

\bibliographystyle{apsrev4-1}
\bibliography{NaK_ref_Arxiv}

\end{document}